\documentclass[aps,pra,twocolumn,showpacs,amsmath,amssymb]{revtex4}

\usepackage{color}
\usepackage{graphicx}

\begin{document}

\title{Dynamically decoupled three-body interactions with applications to interaction-based quantum metrology}

\author{K. W. Mahmud,$^1$ E. Tiesinga,$^1$ and P. R. Johnson$^2$}

\affiliation{$^1$Joint Quantum Institute, National Institute of
Standards and Technology and University of Maryland, 100 Bureau
Drive, Mail Stop 8423, Gaithersburg, Maryland 20899, USA\\
$^2$Department of Physics, American University, Washington, DC
20016, USA}

\begin{abstract}
We propose a stroboscopic method to dynamically decouple the
effects of two-body atom-atom interactions for ultracold atoms,
and realize a system dominated by elastic three-body interactions.
Using this method, we show that it is possible to achieve the
optimal scaling behavior predicted for interaction-based quantum
metrology with three-body interactions. Specifically, we show that
for ultracold atoms quenched in an optical lattice, we can measure
the three-body interaction strength with a precision proportional
to ${\bar n}^{-5/2}$ using homodyne quadrature interferometry, and
${\bar n}^{-7/4}$ using conventional collapse-and-revival
techniques, where ${\bar n}$ is the mean number of atoms per
lattice site. Both precision scalings surpass the nonlinear
scaling of ${\bar n}^{-3/2}$, the best so far achieved or proposed
with a physical system. Our method of achieving a decoupled
three-body interacting system may also have applications in the
creation of exotic three-body states and phases.
\end{abstract}


\maketitle

\emph{Introduction.}$-$ The ultimate precision of a measurement
plays an important role in many areas of physics and technology.
Important examples are gravitational wave detection, atomic
clocks, and magnetometry~\cite{lloyd04}. A current goal of
metrology research is to improve measurement precision using
quantum resources such as squeezing and
entanglement~\cite{lloyd11,linjian13}. Measurement of a parameter
$\gamma$ using an interferometer with ${\bar n}$ independent
particles can achieve a precision limit of $\delta \gamma \propto
{\bar n}^{-1/2}$. This is the shot-noise or standard quantum limit
(SQL), which is the best precision possible for a classical
system. Use of quantum resources can improve this precision to the
Heisenberg limit (HL) $\delta \gamma \propto {\bar
n}^{-1}$~\cite{dowling08}. Recent experiments with photons, atoms
and other systems have achieved sub-shot-noise
precision~\cite{oberthaler10,reidel10}.

Recent analyses of the quantum Cramer-Rao bound with nonlinear
terms in the
Hamiltonian~\cite{luis04,beltran05,boixo07,boixo08,roy08,luis07,rey07,choi08,datta12}
have argued that a precision beyond the Heisenberg limit is
possible. Multibody ($k$-body) interactions can give rise to the
nonlinearity, where its strength $U_k$ corresponds to the
parameter to be estimated. A $k$-body interaction in the
Hamiltonian is predicted to give a scaling of ${\bar n}^{-k}$
using an optimally entangled state and ${\bar n}^{-(k-1/2)}$ even
without entanglement, where ${\bar n}$ is the number of probes.
This so-called ``super-Heisenberg" scaling surpasses the
conventional Heisenberg limit for $k \ge 2$, and reduces to the
SQL and HL for $k=1$ (linear case). Scaling of ${\bar n}^{-3/2}$,
where ${\bar n}$ is the number of photons, has been experimentally
achieved~\cite{napolitano11} in the detection of atomic
magnetization that couples to effective pairwise ($k=2$)
photon-photon interactions. Theory proposals and analysis exist on
performing interaction-based metrology with two-body interactions
in a number of
systems~\cite{boixo08,boixo09,tiesinga13,napolitano10,sewell13,tilma10,wolley08,chase09}.
Although there continues to be debate on the correct way to count
resources for nonlinear metrologies~\cite{lloyd11,lloyd12}, the
potential for either enhanced or new types of measurement
exploiting particle-particle interactions in quantum systems
deserves further investigation.

In this paper, we propose a method for achieving the optimal
precision scaling for an interaction-based quantum metrology
exploiting three-body interactions ($k=3$), within an
experimentally realizable physical system. For ultracold atoms in
an optical lattice, we show that the elastic three-body
interaction strength can be measured with a precision scaling of
${\bar n}^{-5/2}$ using a quadrature method~\cite{tiesinga13} and
${\bar n}^{-7/4}$ using conventional collapse and revival
techniques~\cite{will10,porto07,johnson11,mahmud13a,mahmud13b,mahmud14a},
where ${\bar n}$ is the average number of atoms per site. These
precision limits surpass the interaction-based scaling of ${\bar
n}^{-3/2}$, the best possible scaling so far
realized~\cite{napolitano11} or proposed~\cite{boixo08,tiesinga13}
with a physical system. Our analysis and results add to the
toolbox of quantum metrology and may find applications in
demanding precision measurements.

Ultracold atoms in a shallow optical lattice, when quenched to a
deep lattice, exhibit matter-wave collapse and revivals with
signatures of multibody interactions~\cite{will10}. The challenge
for exploiting the three-body physics is to effectively turn off
or decouple the, typically stronger, influence of the two-body
interactions on the dynamics. We propose achieving this via a
dynamical decoupling protocol in which a Feshbach resonance is
used to switch the sign of $U_2$ periodically while $U_3$ is
unchanged. This cancels the influence of two-body interactions on
the dynamics, decoupling the three-body physics in stroboscopic
measurements. This technique for decoupling two- and three-body
interactions may also have application to generating novel
three-body states with topological characteristics, such as the
Pfaffian state~\cite{greiter91} and other exotic phases and
phenomena~\cite{hafezi13,daley14,daley09,mazza10}. We note that
one can also modify the multibody interactions by dressing atoms
using microwave or radio-frequency
radiation~\cite{petrov14a,petrov14b}.

\emph{State preparation.}$-$ Our system is an ultracold gas of
bosons in an optical lattice, which can initially be described by
the single-band Bose-Hubbard Hamiltonian,
\begin{eqnarray}
H_i=-J_i\sum_{\langle jj' \rangle}\left( b_{j}^{\dagger}b_{j'}+
{\rm h.c.} \right) +
\frac{U_{2,i}}{2}\sum_{j}n_{j}\left(n_{j}-1\right),
\label{eqn:boseHubb}
\end{eqnarray}
where $j, j'$ are indices to lattice sites, $n_j=
b_{j}^{\dagger}b_{j}$, $J_i$ is the hopping parameter, $U_{2,i}$
is the initial on-site two-body atom-atom interaction strength,
and only nearest-neighbor tunneling is assumed. The Hamiltonian
holds for 1D, 2D, and 3D systems. The total number of particles is
$N=\sum_{j} \langle b_{j}^{\dagger}b_{j}\rangle$, where ${\bar
n}=N/M$ is the mean occupation per site and $M$ is the number of
sites.

\begin{figure}[t]
\vspace{-0.0cm}
\begin{center}
  \includegraphics[width=0.45\textwidth,angle=0]{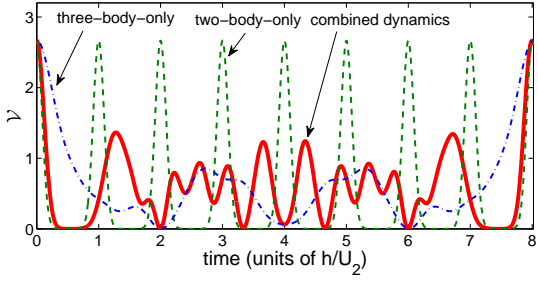}
\end{center}
\vspace{-0.7cm} \caption{\label{fig:u2u3dynamics} (color online).
Visibility as a function of time showing the effects of two- and
three-body interactions in the quench dynamics of ultracold atoms
in an optical lattice, where ${\bar n}=2.67$. The solid line (red)
is for the combined $U_2$ and $U_3$ dynamics, while the dashed and
dot-dashed lines are for $U_2$-only and $U_3$-only dynamics,
respectively. To extract the three-body scaling, we need to
isolate the $U_3$-only dynamics from the combined time trace.}
\end{figure}

We prepare our initial state as a superfluid in a shallow lattice,
which in the limit of $U_{2,i}/J_i \to 0$ approaches a product of
coherent states, one at each lattice site. We then suddenly
increase (quench) the depth of the optical lattice such that
tunneling is suppressed~\cite{will10}. The effective Hamiltonian
for the post-quench dynamics is
\begin{eqnarray}
H_f=\frac{U_{2}}{2}\sum_{j}b_j^{\dagger}b_j^{\dagger} b_j
b_j+\frac{U_{3}}{6}\sum_{j}b_j^{\dagger}b_j^{\dagger}b_j^{\dagger}
b_j b_j b_j+\mathcal{O}(U^3_{2}),
\end{eqnarray}
where $U_{2}$ and $U_{3}$ are, respectively, the effective two-
and three-body interaction strengths in the deep lattice. The
effective three-body interaction arises due to collision-induced
virtual excitations to higher bands or vibrational levels of the
isolated sites~\cite{johnson09}. Approximating the bottom of the
deep lattice as an isotropic harmonic potential with frequency
$\omega_f$, $U_{3}$ is attractive and given
by~\cite{johnson09,johnson12}
\begin{equation}
U_{3}=-c_f U^2_{2}/(\hbar \omega_f)+\mathcal{O}(U^3_{2}),
\end{equation}
with $c_f=1.344$. There exist additional higher-body corrections
of order $\mathcal{O}(U^3_{2})$ whose strengths are much
smaller~\cite{johnson09}, and that are omitted in this article.

\begin{figure}[t]
\vspace{-0.0cm}
\begin{center}
  \includegraphics[width=0.45\textwidth,angle=0]{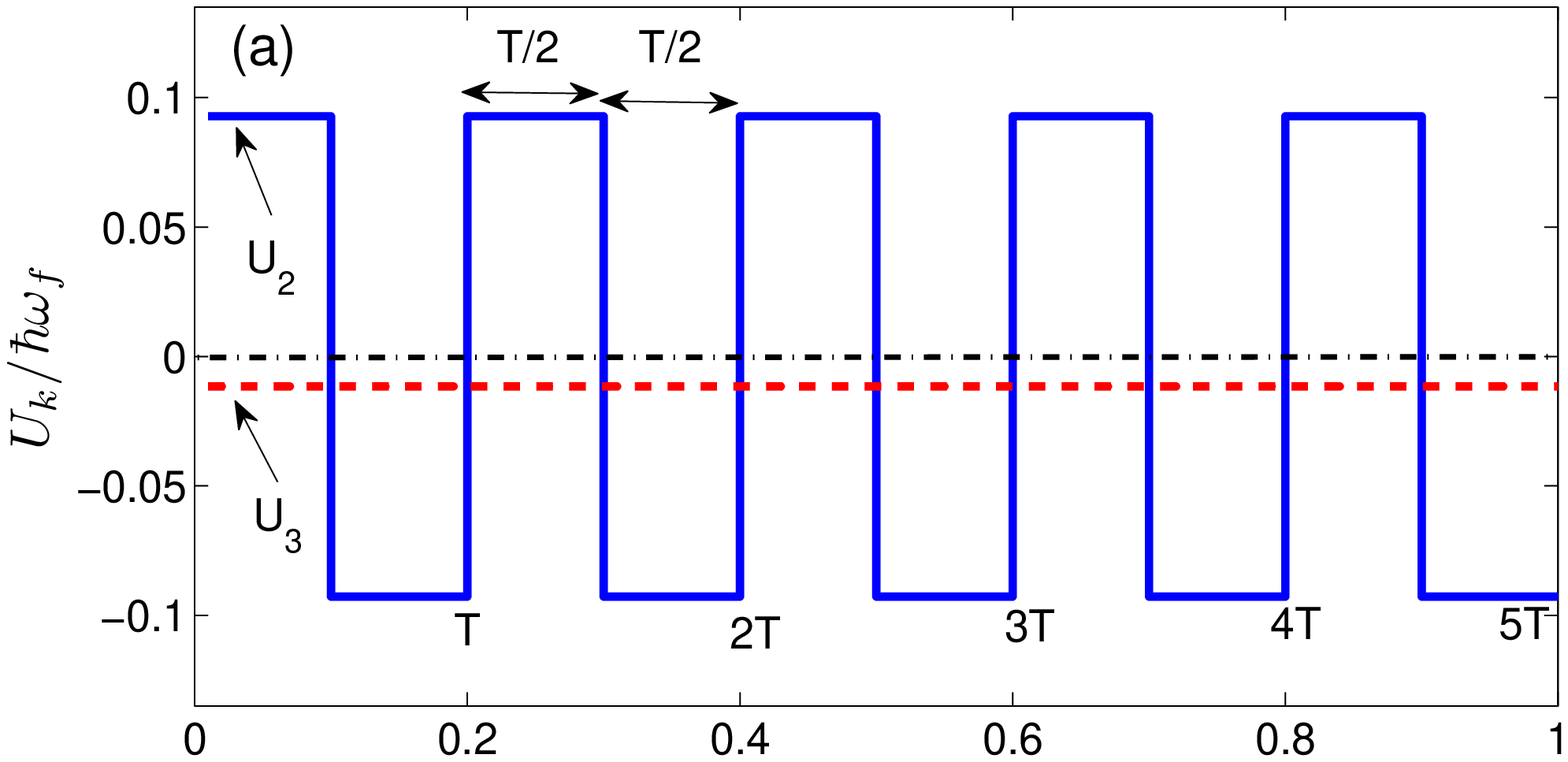}
\end{center}
\vspace{-0.9cm}
\begin{center}
  \includegraphics[width=0.45\textwidth,angle=0]{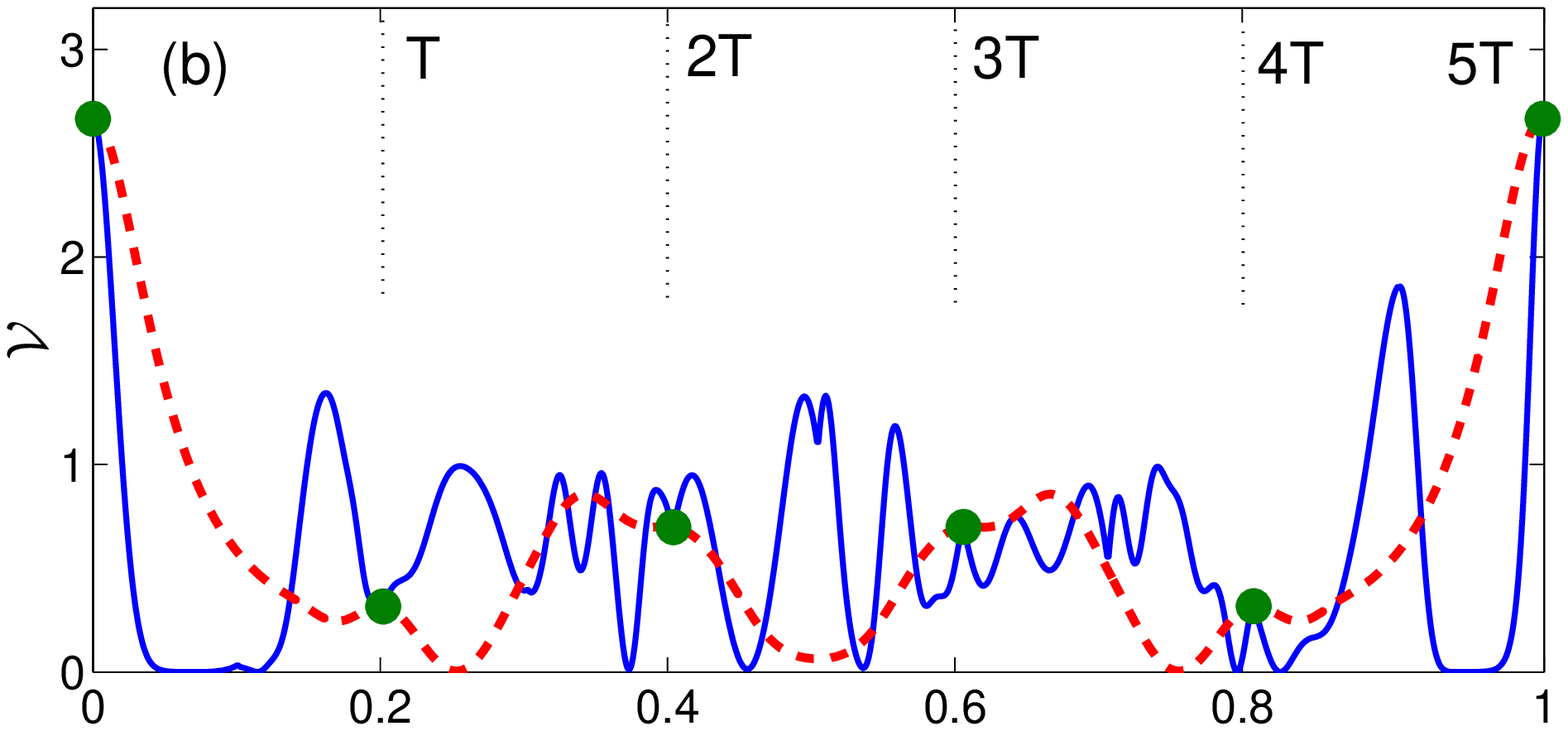}
\end{center}
\vspace{-1.1cm}
\begin{center}
  \includegraphics[width=0.45\textwidth,angle=0]{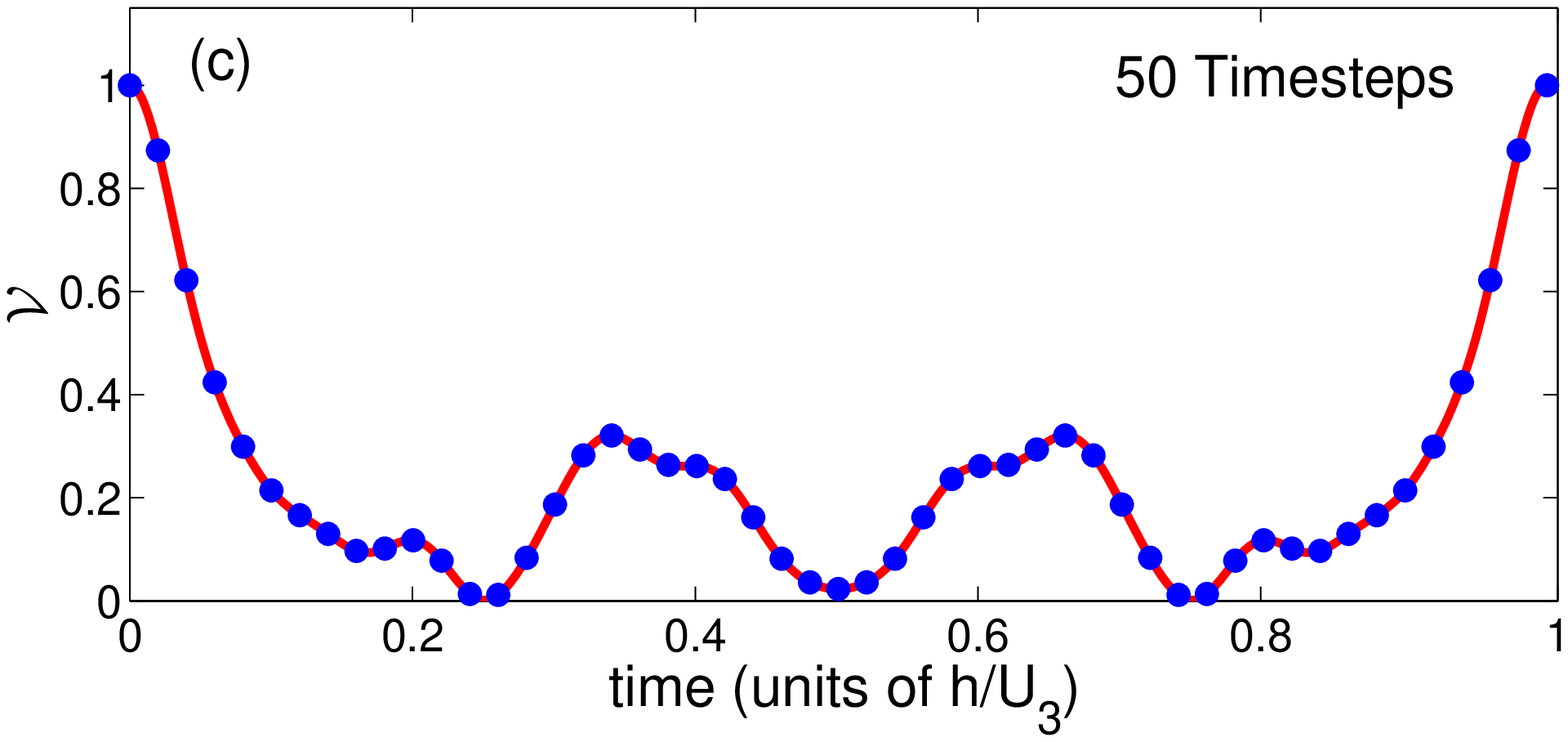}
\end{center}
\vspace{-0.7cm} \caption{\label{fig:protocol} (color online). A
schematic of the dynamical decoupling protocol to effectively turn
off the two-body interactions and isolate the dynamics due to the
three-body interaction. Panel (a) shows the switching of the sign
of $U_2$ as a function of time, which can be done through the use
of a Feshbach resonance. The three-body strength $U_3$ remains
constant and negative. Panel (b) shows the visibility (solid line)
as a function of time under the influence of the periodically
changing $U_2$ shown in Panel (a), and ${\bar n=2.67}$. Also shown
as a dashed line is the time evolution due to only $U_3$. The two
curves intersect after each interval of duration $T$, indicated by
the green circles. Panel (c) shows the intersection points for a
time interval $T$ that is 10 times smaller, coinciding even more
with the pure three-body time trace. The solid line shows the
dynamics for $U_3$-only evolution and markers correspond to
stroboscopic measurements at multiples of period $T$.}
\end{figure}

The initial superfluid state is not an eigenstate of the deep
lattice. Since the initial state is separable and the lattice
sites are decoupled after the quench, the state at each site
evolves as $|\Psi(t) \rangle=\sum_n c_n e^{-i E_n t/\hbar} |n
\rangle$, with Fock states $|n \rangle$ containing $n=0,1,2,...$
atoms in the lowest vibrational state of a lattice site. The
initial amplitudes $c_n$ are given for a coherent state, and the
energies are $E_n=U_2 n(n-1)/2+U_3 n(n-1)(n-2)/6$; $\hbar=h/2\pi$
is the reduced Planck's constant. After a hold-time $t$ in the
deep lattice, the lattice is turned off and the atoms expand
freely. Absorption imaging of the atomic spatial density yields
the quasi-momentum distribution $n_{k}(t)
=(1/M)\sum_{j,j'}e^{ik(j-j')}\rho_{jj'}(t)$, where
$\rho_{jj'}(t)=\langle b_{j}^{\dagger}b_{j'}\rangle$ and $k$ is
the lattice wavevector. A measurement of the normalized observable
${\cal V}(t)=n_{k=0}(t)/M$ shows collapse and revival oscillations
driven by the two- and three-body
interactions~\cite{will10,johnson09,johnson12}.

The solid curve in Fig. 1 shows a representative dynamics for the
experimentally relevant values of $U_{2}=0.0928 \hbar \omega_f$
and ${\bar n}=2.67$. The effective three-body interaction
strength, using Eq.~3, is $U_{3}=-0.1247 U_{2}$. We see a complex
pattern of oscillations: the faster oscillations are caused by
$U_{2}$, modified by a slower envelope due to $U_{3}$. Will {\it
et al.}~\cite{will10} has observed these predicted visibility
oscillations and demonstrated the presence of multibody
interactions by analyzing the oscillation frequencies. The dashed
and dot-dashed lines show the simpler behavior that would result
from pure $U_2$-only and $U_3$-only dynamics, respectively.
Reference~\cite{tiesinga13} showed that the two-body atom-atom
interaction strength $U_{2}$ can be extracted from the visibility
with a minimal possible uncertainty scaling as $\bar{n}^{-3/2}$,
when the measurement is optimized using a quadrature
interferometry method, and $\bar{n}^{-3/4}$ without optimization.

\emph{Dynamical decoupling protocol.}$-$ We propose a protocol
similar to dynamical decoupling~\cite{koschorreck10} or spin-echo
methods. Our approach is based on the key observation that in Eq.
3 the value of $U_{3}$ is independent of the sign of $U_{2}$. This
allows one to change $U_{2}$ to $-U_{2}$ using an external
magnetic field near a collisional Feshbach
resonance~\cite{tiesinga10}, without changing the value of $U_3$.
Specifically, we average out the influence of two-body
interactions by alternating between interaction strength set to
$|U_{2}|$, for a time interval $T/2$, and then switching to
$-|U_{2}|$, for the next $T/2$ time interval, thus completing one
full time step of duration $T$. Figure 2(a) shows a schematic of
the protocol. As the $k$-body interaction terms commute, we can
write the dynamics in one time-step $T$ in terms of time-ordered
unitary evolution operators, giving
\begin{eqnarray}
|\Psi(t+mT)\rangle =
               {\hat{\cal U}}_3(U_3,t){\hat{\cal
               U}}_2(|U_2|,t)|\Psi(mT)\rangle
\end{eqnarray}
for $mT<t<mT+T/2$ where $m=0,1,2..$, and
\begin{eqnarray}
|\Psi(t+mT+T/2)\rangle=&& {\hat{\cal U}}_3(U_3,t){\hat{\cal
  U}}_2(-|U_2|,t) \nonumber \\
  &&\times |\Psi(mT+T/2)\rangle
\end{eqnarray}
for $mT+T/2<t<(m+1)T$. Here ${\hat{\cal U}}_2(U_2,t)=e^{-i U_2
b^{\dagger}b^{\dagger} b b t/(2 \hbar)}$ and ${\hat{\cal
U}}_3(U_3,t)=e^{-i U_3 b^{\dagger}b^{\dagger}b^{\dagger} b b b
t/(6 \hbar)}$. Some algebra shows that at $\tau=T$,
$|\Psi(t+T)\rangle=e^{-i U_3 b^{\dagger}b^{\dagger}b^{\dagger} b b
b T/(6\hbar)}|\Psi(t)\rangle$; the dynamics due to 2-body
interactions cancel exactly at the end of each period $T$, leaving
only the $U_3$ contribution.

Note that for any other time $t$, however, both $U_2$ and $U_3$
influence the dynamics, giving the complex dynamics depicted in
Fig. 2(b). Only for times that are integer multiples of $T$ does
the combined evolution yield a visibility that corresponds to the
$U_3$-only time trace. By making the period $T$ smaller we can
obtain a nearly continuous sampling of the $U_3$-only dynamics, as
depicted in Fig.~2(c). In other words, the protocol gives
$|\Psi(mT)\rangle= {\hat{\cal U}}_3(U_3,T)^m |\Psi(t=0)\rangle$,
for integer $m$. As an aside, we note that in our method not only
two-body but all even-body interactions, such as the effective
four-body interaction, are approximately cancelled in the dynamics
because their leading-order dependence on $U_2$ is even.

\begin{figure}[t]
\vspace{-0.0cm}
\begin{center}
  \includegraphics[width=0.4\textwidth,angle=0]{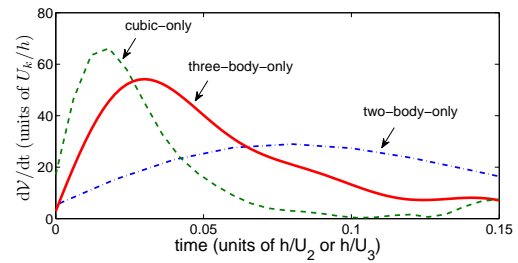}
\end{center}
\vspace{-0.7cm} \caption{\label{fig:scaling} (color online). The
derivative of the visibility ${\cal V}$ as a function of time for
two-body-only (dot-dashed line), three-body-only (solid), and
cubic nonlinearity (dashed) simulations. We use ${\bar n}=2.67$.}
\end{figure}

\emph{Nonlinear metrology for three-body interactions.}$-$ To
optimally extract the interaction strength $U_3$ from the
visibility, we need to minimize the fractional uncertainty
\begin{equation}
\delta U_3/U_3 \propto \frac{\Delta {\cal V}}{|d {\cal V}/dt|},
\end{equation}
obtained by error propagation; $\Delta {\cal V}$ is the
uncertainty of the visibility, which we obtain from the variance
of $n_{k=0}$. Optimal sensitivity is determined by a trade-off
between maximizing the derivative and minimizing the uncertainty
in the visibility.

Figure 3 shows a comparison of the time derivative of the
visibility for two-body-only and three-body-only dynamics. To
facilitate comparison, both the $x$ and $y$ axes have been scaled
to natural units of time $h/U_k$, where $k=2$ or $3$ for the
two-body-only or three-body-only simulations, respectively. For
small times the visibility time trace is steeper for the
$U_3$-only simulation, in comparison to the $U_2$-only simulation.
In fact, for larger values of ${\bar n}$ (not shown here), the
difference in slope becomes even more pronounced. For $U_2$-only
dynamics, there exists an analytic expression for the visibility,
${\cal V}= {\bar n} e^{2 {\bar n} (\cos(U_2t/\hbar)-1)}$, and its
variance, for an initial coherent state. Although no closed-form
expression exists for $U_3$-only dynamics, for short times and an
initial coherent state we can derive a semi-analytic series
expansion. Using a combination of analytics and numerics, we find
the precision scaling from collapse and revival measurements of
${\cal V}$, given by
\begin{equation}
\delta U_3/U_3 \propto M^{-1/2} {\bar n}^{-7/4}.
\end{equation}
The scaling with ${\bar n}$ of $7/4$ is greater than $3/2$, the
best so far proposed or achieved exploiting nonlinear
interactions~\cite{napolitano11}.

\begin{figure}[t]
\vspace{-0.0cm}
\begin{center}
  \includegraphics[width=0.4\textwidth,angle=0]{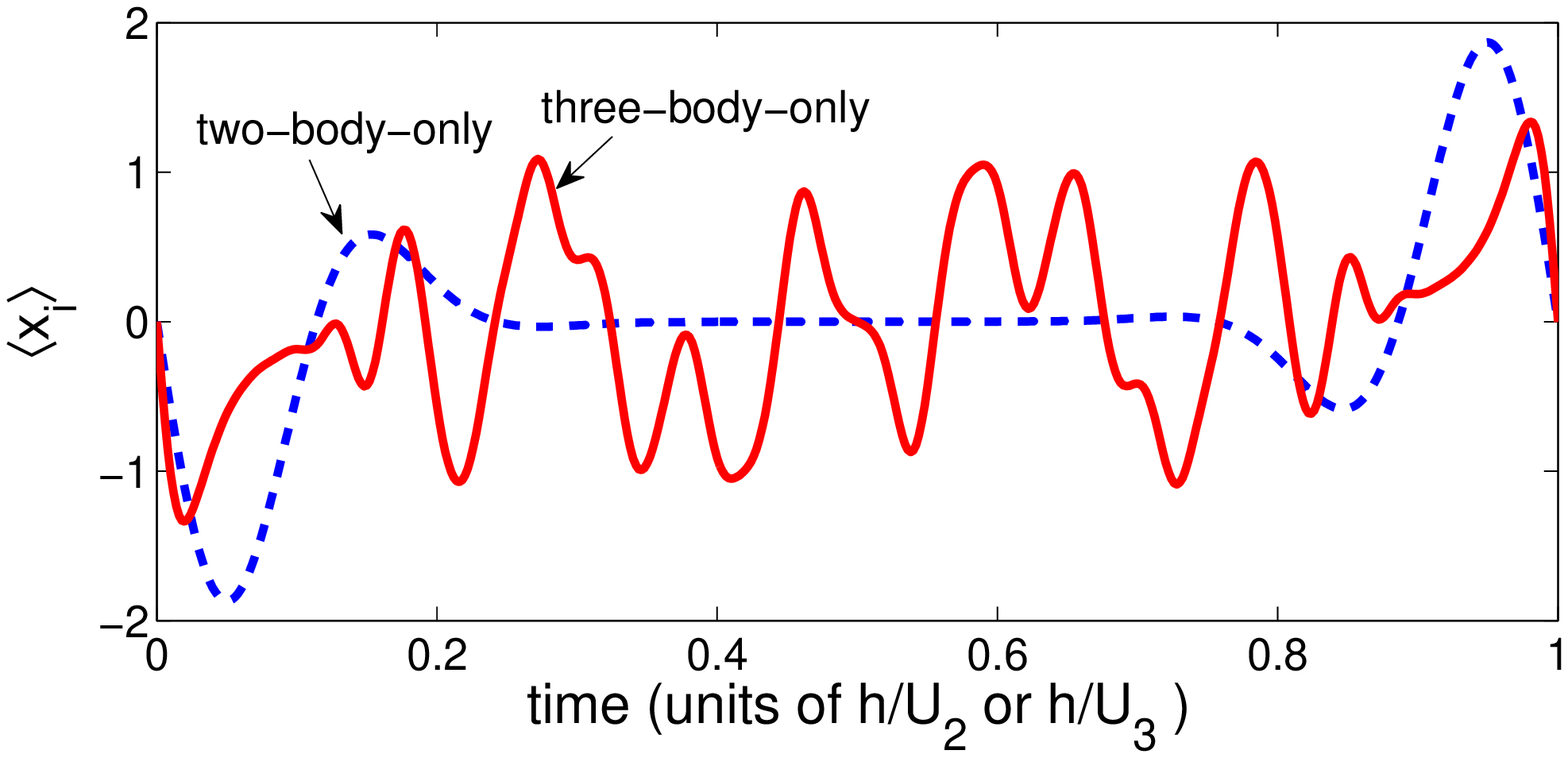}
\end{center}
\vspace{-0.9cm}
\begin{center}
  \includegraphics[width=0.4\textwidth,angle=0]{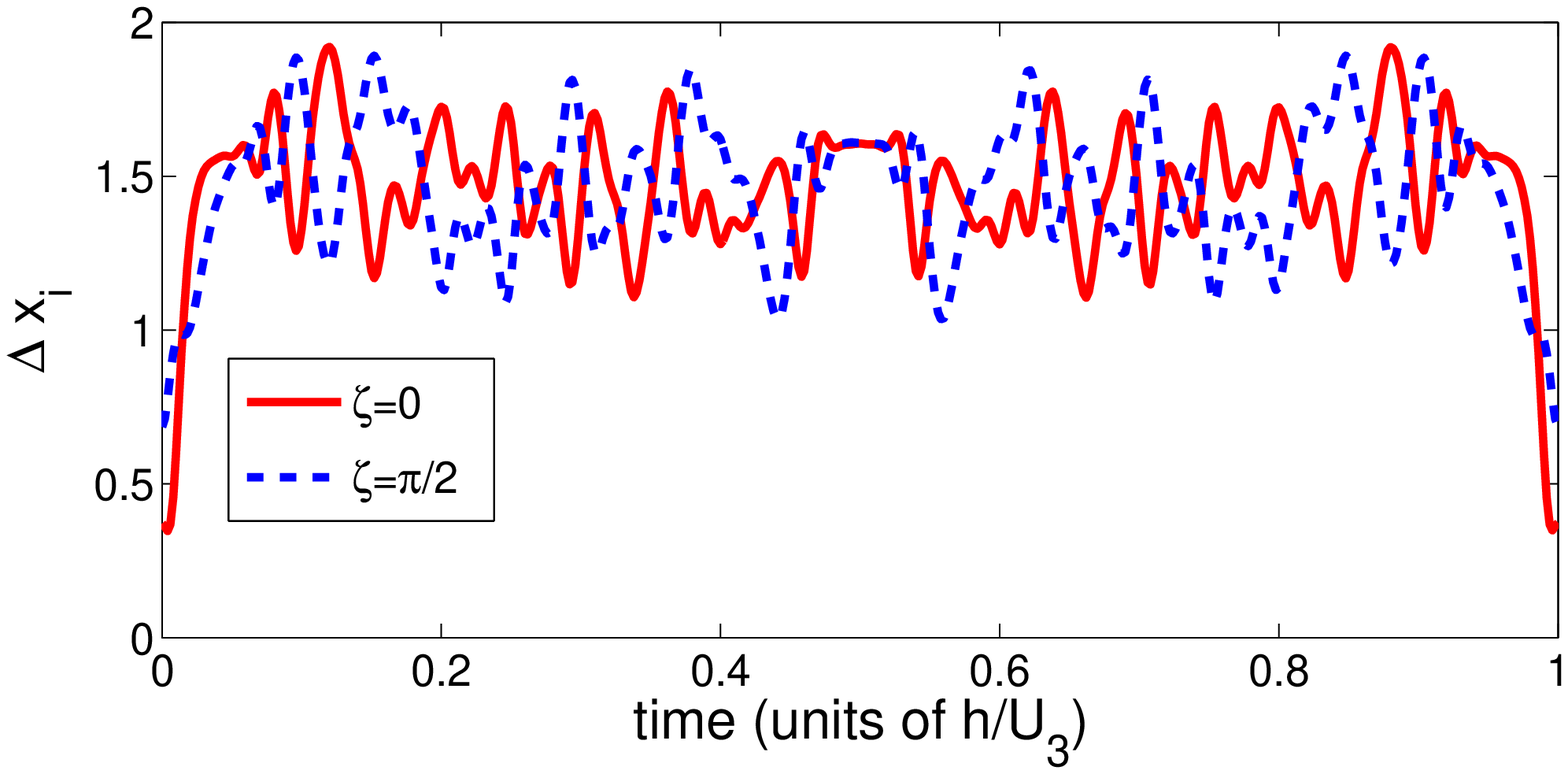}
\end{center}
\vspace{-0.7cm} \caption{\label{fig:u2u3dynamics} (color online).
Panel (a) quadrature dynamics $\langle x_i \rangle$ as a function
of time for two-body-only and three-body-only interactions for
${\bar n}=4.68$ and $\zeta=0$. Panel (b) shows the evolution of
quadrature fluctuations for three-body-only interactions for
$\zeta=0,\pi/2$ and ${\bar n}=4.68$.}
\end{figure}

The scaling is valid in the limit ${\bar n} \gg 3$, where ${\bar
n}({\bar n}-1)({\bar n}-2)={\bar n}^3-3{\bar n}^2+2{\bar n}$ is to
a good approximation ${\bar n}^3$ and the quadratic correction is
negligible. Figure 3 also shows a simulation of the derivative of
the visibility when the three-body interaction Hamiltonian is
replaced by the cubic nonlinear Hamiltonian $U_3 (b^{\dagger}
b)^3/6$. This shows that for ${\bar n}=2.67$ we have not reached
the asymptotic regime of large ${\bar n}$. We note that $\delta
U_3/U_3$ scales as $M^{-1/2}$, the standard quantum limit in the
number of lattice sites. This is expected as the initial state is
site separable and, furthermore, tunneling is turned off after the
quench. It is as if we have performed $M$ independent
measurements. Consequently, the overall scaling in total particle
number $N={\bar n}M$ is sub-Heisenberg limit. Nevertheless, we
find that the improved scaling in ${\bar n}$ promises real
improvements in measurement precision.

Measurement theory suggests that for a cubic nonlinearity the best
possible precision scaling is ${\bar n}^{-3}$ using entangled
states and ${\bar n}^{-5/2}$ using product states. Equation 7
shows that a measurement of visibility in standard collapse and
revival experiments does not give the optimal scaling. However, we
find that the quadrature interferometry method, details of which
are given in Ref.~\cite{tiesinga13}, can be used to further
improve the scaling behavior. Using that method, we measure the
time evolution of the field quadratures, $X_{k=0}=1/{\sqrt
M}\sum_{i} x_i$, where $x_i=(e^{-i \zeta} b_i+e^{i \zeta}
b^{\dagger}_i)/2$ and $\zeta$ is a controllable phase. We then
have, $\delta U_3/U_3 \propto M^{-1/2} \Delta x_i/|d\langle x_i
\rangle/dt|$~\cite{tiesinga13}, with $\Delta
x_i=\sqrt{\langle(x_i-\langle x_i \rangle)^2\rangle}$, and we can
optimize with respect to $\zeta$.

Figure 4(a) shows the dynamics of the quadrature $\langle x_i
\rangle$ for two-body-only and three-body-only cases. In comparing
the two traces, we see that initially the derivative of the
three-body-only case is steeper by a factor of ${\bar n}$. Figure
4(b) shows the variance $\Delta x_i$ for three-body-only
simulations for different phases $\zeta$. It is smallest for small
times and integer multiples of $h/U_3$. We choose the phase
variable such that the numerator does not degrade the scaling
enhancements gained from the steepness of the time trace. This
occurs for $\zeta=\pi/2$, and we find that the best possible
scaling is
\begin{equation}
\delta U_3/U_3 \propto M^{-1/2} {\bar n}^{-5/2}.
\end{equation}
This gives the optimal precision for three-body interactions for
an input state that is not entangled. This scaling improves upon
the ${\bar n}^{-7/4}$ scaling of Eq. 6.

In principle, we can generalize beyond the use of an initial
coherent state, and consider number squeezed initial states (these
can be described using the Gutzwiller
approximation~\cite{jaksch98}). Alternative initial states do not
affect the dynamical decoupling protocol, and we still obtain
stroboscopic evolution under $U_3$-only, however, the scaling with
${\bar n}$ will degrade since the measurements are sensitive to
phase fluctuations.

\emph{Implementation challenges.}$-$ Our method promises improved
measurement of $U_3$ even for modest ${\bar n}$. Recently, Will
{\it et al}~\cite{will10} obtained $\delta U_3/U_3=30 \%$ with
$N=2 \times 10^5$ and ${\bar n}= 2.5$ and Ma {\it et
al}~\cite{ma11} obtained $\delta U_3/U_3=5 \%$. For our proposed
method and with ${\bar n}=4.5$, for example, it is possible in
principle to improve the precision by a factor of 6 with the
collapse and revivals and 20 with the quadrature method. To
achieve the predictions of nonlinear scaling in this paper,
experiments need to minimize uncertainties that originate from
fluctuations in the total atom number, lattice depth fluctuations
from lasers, residual tunneling, and errors in the pulse sequence.
Other factors such as three-body recombination
losses~\cite{tiesinga10}, effective range
corrections~\cite{johnson12}, optical lattice inhomogeneities and
the higher-order corrections to the multibody effective
interactions~\cite{johnson12} may also need to be considered.

\emph{Conclusion.}$-$ We propose a dynamical decoupling method to
average-out the influence of two-body (and higher even-body)
interactions in ultracold atom dynamics. Our method for achieving
a system dominated by three-body interactions should have a number
of applications, including possible realization of novel
three-body phases and states. In this paper, we describe how to
achieve nonlinear quantum metrology scaling for three-body
interactions with an experimentally realizable physical system. We
predict a scaling of ${\bar n}^{-5/2}$ using a quadrature method
and ${\bar n}^{-7/4}$ in collapse and revivals of momentum
distribution. These results are a significant improvement over any
scaling so far experimentally achieved or proposed with a physical
system exploiting particle-particle interactions.

\emph{Acknowledgments.}$-$ We acknowledge support from the US Army
Research Office under Contract No. 60661PH.

\end{document}